\documentclass[pdftex,prb,aps,twocolumn,showpacs,superscriptaddress]{revtex4}
\usepackage[pdftex]{graphicx}
\usepackage{latexsym}
\usepackage{amsmath}
\begin{document}

\title{Angle dependent quasiparticle weights in correlated metals}

\author{Pouyan Ghaemi}
\affiliation{Department of Physics, Massachusetts Institute of
Technology, Cambridge, Massachusetts 02139}

\author{T. Senthil}
\affiliation{Department of Physics, Massachusetts Institute of
Technology, Cambridge, Massachusetts 02139}

\author{P. Coleman}
\affiliation{Department of Physics and Astronomy, Rutgers University, Piscataway, NJ 08854-8019}

\date{\today}
\begin{abstract}
The variation in the quasiparticle weight $Z$ on moving around the
fermi surface in correlated metals is studied theoretically. Our
primary example is a heavy Fermi liquid treated within the standard
hybridization mean field theory. The most dramatic variation in the
quasiparticle weight happens in situations where the hybridization
vanishes along certain directions in momentum space. Such a
``hybridization node" is demonstrated for a simplified model of a
Cerium-based cubic heavy electron metal.  We show that the
quasiparticle weight varies from almost unity in some directions, to
values approaching zero in others.  This is accompanied by a similar
variation in the quasiparticle effective mass. Some consequences of
such hybridization nodes and the associated angle dependence are
explored. Comparisons to somewhat similar phenomena in the normal
metallic state of the cuprate materials are discussed. A
phenomenological picture of the pseudogap state in the cuprates with a
large Fermi surface with a severely anisotropic spectral weight is explored.

\end{abstract}
\newcommand{\be}{\begin{equation}}
\newcommand{\ee}{\end{equation}}
\maketitle

\section{Introduction} The normal state of the cuprate materials is
often described (at least empirically) as a non-fermi liquid metal.  A
remarkable feature of this metal is the presence of significant
momentum space anisotropy\cite{dicho}: the extent to which Fermi
liquid theory {\em fails} depends strikingly on which part of a
nominal Fermi surface is being probed. In optimally doped systems the
quasiparticle-like peaks measured in photoemission experiments are
typically much broader along the `antinodal' direction near the edges
of the Brillouin zone than along the diagonal `nodal' direction. The
difference is even more striking in the underdoped cuprates where a
pseudogap opens - apparently only near the antinodal regions leaving
behind a gapless `Fermi arc' centered at four nodal
points\cite{norman}. Somewhat similar phenomena have been reported
even in the overdoped cuprates based on transport experiments though
the anisotropy weakens with increasing doping\cite{hussey}.

Theoretical understanding of such phenomena in the cuprates is
primitive and is hampered by the lack of a suitable framework for
describing non-fermi liquid phenomena\cite{dmft}. However the cuprates are but one example of
a host of correlated metals that have been studied over the
years. Fermi liquid theory does not always fail in such metals.
Motivated by the observed momentum space anisotropy in the cuprates we
therefore pose the opposite general question: does the extent to which
Fermi liquid theory {\em work} depend dramatically on where one is on
the Fermi surface in a correlated Fermi liquid metal? As there is a
firm theoretical framework to discuss Fermi liquid metals, this
question can be expected to yield more easily to progress.

The most celebrated success of Fermi liquid theory is provided by the
`heavy Fermi liquid' state of rare earth alloys.  These have
quasiparticle effective masses as high as 100-1000 times the bare
electron mass and an associated small quasiparticle weight $Z$ at the
Fermi surface\cite{colrev}. The main purpose of the present paper is
to discuss the variation in the quasiparticle weight $Z$ on moving
around the Fermi surface. Indeed $Z$ is a convenient measure of the
extent to which Fermi liquid theory works in a Fermi liquid. The
theoretical approach we use is the standard hybridization mean field
theory for Kondo lattice models of the rare earth alloy. The variation
of $Z$ may be linked to the internal orbital structure of the Kondo
singlet that forms between the local moments and the conduction
electrons. This internal orbital structure derives from the symmetries
of the atomic orbital occupied by the local moment and the conduction
electron band it is coupled to. In the hybridization mean field theory
this leads to angle dependence of the hybridization on going around
the Fermi surface. The most dramatic variation occurs when the
hybridization vanishes along some directions. Along such hybridization
nodes $Z \sim o(1)$ but can become very close to zero along other
directions. We demonstrate the possibility of such hybridization nodes
in a simplified model appropriate for a $Ce$-based cubic
system. Recent angle resolved photoemission experiments\cite{photoexp}
have begun to probe the structure of the electronic excitations of the
heavy Fermi liquid. We expect that the physics described in this paper
may be probed in the near future.

Inspired by these calculations appropriate to heavy electron systems,
we consider the possibility that the pseudogap regime of the
underdoped cuprates may actually have a large band-structure Fermi
surface but with strongly angle dependent $Z$. Several experimental
results on the underdoped cuprates are examined in this light. Such a
pseudogap state has some attractive phenomenological features - in
particular it provides one possible reconciliation between recent high
field quantum oscillation experiments\cite{taill1,harrison} and older
ARPES reports of gapless `Fermi arcs'. However such a large fermi
surface Fermi liquid state also has a number of problems with other
experiments making it unappealing as a serious theory of the
underdoped cuprates. A {\em non-fermi liquid} version of such a large
Fermi surface state might perhaps resolve these difficulties but
theoretical description of such a state remains out of reach.

\section{Kondo singlets with internal orbital structure} The heavy
fermion materials are conveniently modeled as Kondo lattices, {\em
i.e.} a periodic lattice of local moments coupled by magnetic exchange
to a separate band of conduction electrons\cite{don}. At low
temperatures the local moments are absorbed into the Fermi sea of the
metal through Kondo singlet formation. In a typical heavy electron
metal the local moments occupy atomic $f$-orbitals. The conduction
electrons derive from bands with different symmetry ($s$, $p$ or
$d$). The Kondo singlet that forms between a local moment and a
conduction electron will therefore have nontrivial internal orbital
structure.  In the low temperature heavy fermi liquid phase this
orbital structure leads to pronounced anisotropies between various
parts of the Fermi surface. A close analogy is with the physics of
unconventional superconductors where Cooper pairs with non-trivial
internal orbital structure condense leading to anisotropic
superconductivity.  In the heavy Fermi liquid case such anisotropic
effective masses are known to occur and have been discussed
theoretically using a renormalized band theory approach\cite{zwi}.

In the present paper we will mainly focus on the quasiparticle
spectral weight $Z$ which is a measure of the extent to which Fermi
liquid theory works. To illustrate our point we focus specifically on
{\em Ce} based heavy electron materials with the {\em Ce} ion in a
$f^1$ state\cite{cox}. We also assume cubic symmetry. Such a $Ce$ ion
has, after considering the effect of spin-orbit coupling and crystal
field splitting, a low energy Kramers doublet that couples to a
separate conduction band. We treat the corresponding Kondo lattice
model within the slave boson mean field
approach\cite{read,millee}. This approach is particularly well suited
to describing the heavy Fermi liquid phase. At the mean field level
there are two bands - one derived from the f-moments and the other
from the conduction electrons - that are hybridized. Physically the
hybridization amplitude is a measure of the Kondo singlet formation.
We show that this amplitude has strong momentum dependence coming from
the symmetry of the f-orbital. Thus the true quasiparticles at the
fermi surface are angle dependent admixtures of the $f$-fermions and
the conduction electrons. Most remarkably we show that our simplified
model naturally has directions where the hybridization vanishes. These
hybridization nodes have a number of consequences.  Most importantly
it leads to a fermi surface structure where along the hybridization
nodes the true (large) Fermi surface is contained within the original
small Fermi surface of the conduction electrons. Thus along these
directions the true quasiparticle mostly has $c$-character with weak
admixture to $f$. Along other directions the situation is
reversed. Now the physical electron spectral weight depends on the
extent to which the conduction electron contributes to the
quasiparticle state of the true large Fermi surface. This then leads
to the dramatic variation of the quasiparticle weight discussed in the
Introduction.

In passing we note that hybridization nodes have previously been proposed in the context of theories of gapless Kondo {\em insulators}\cite{ikeda,moreno}. When present such nodes have rather different effects in metallic heavy electron systems as we discuss below. Hybridization nodes are also present in toy Kondo lattice models where each local moment is coupled to a conduction electron at a neighboring site\cite{pouyan}. Though such models are not directly relevant to heavy electron systems they capture some of the same physics described in this paper.

\section{Anderson Model for a Cerium ion}\label{ander}
 We begin by briefly reviewing the Anderson model describing a Cerium $f^1$ impurity in a metallic host.
The $f$ states have orbital angular momentum $l = 3$ so that on including the spin  there are $2\ (2\times 3+1)=14$
quantum states in this orbital. Spin-orbit coupling breaks the
degeneracy of this orbital into two sets of states with $J=7/2$ and
$J=5/2$ where $J$ is the total angular momentum
$\left(J=s+l\right)$. The $J=5/2$ states have lower energy and so we will
concentrate on them. In a cubic environment
crystal fields will further split the $J=5/2$ states into a doublet
(lower energy) and a quadruplet (higher energy) states. We will
concentrate on the lower energy doublet described by \cite{cox}:
\begin{eqnarray}\label{egn}
|M = 1> = \
&&\left(\frac{1}{6}\right)^{1/2}|J_z=-\frac{5}{2}\rangle-\left(\frac{5}{6}\right)^{1/2}|J_z=\frac{3}{2}\rangle,
\nonumber\\ |M = 2> = \ &&\
\left(\frac{1}{6}\right)^{1/2}|J_z=\frac{5}{2}\rangle-\left(\frac{5}{6}\right)^{1/2}|J_z=-\frac{3}{2}\rangle.
\end{eqnarray}
As expected these states go into each other under time reversal. Now
consider coupling of this doublet to a band of conduction electrons
$c_{\textbf{k} \sigma}$. We assume that the $f$-electron in a state
$M$ can hybridize with the appropriate partial wave of the
$c$-electron also in the same state $M$. The coupling may therefore
be modeled by the Anderson impurity Hamiltonian\cite{anderson}:
\begin{equation} \label{ande}
\begin{split}
H=& \sum_{k,M} \varepsilon_\textbf{k} c^\dagger_\textbf{k}
c_\textbf{k}+\mu_f \sum_M f^\dagger_M f_M + U\sum_{M,M'} n_M n_{M'} \\
& \sum_{k,M} V_k c_{k M}^\dagger f_M+V^*_k f^\dagger_M c_{k M},
\end{split}
\end{equation}
with the electron partial wave operator $c_{k,M}$ defined through
\begin{equation}\label{base}
c_{k,M}^\dagger
=\sum_{\sigma}\int
\frac{d\Omega_{\hat  {\bf k}}}{4 \pi}
c_{\textbf{k},\sigma}^\dagger \langle \textbf{k}, \sigma | k,M
\rangle,
\end{equation}
where the integral is taken over all
directions of the vector $\hat \bf{k}$. For simplicity we
assume further that $V_k = V$ independent of $\textbf{k}$.

Focussing now on the strong correlation limit of large $U$ we
restrict the $f$-occupation to be one, {\em i.e.} we impose the
constraint
\begin{equation}\label{focc}
\sum_M f^{\dagger}_M f_M = 1.
\end{equation}
The standard Schrieffer-Wolf transformation\cite{schw,schc} then gives the ``Kondo" effective Hamiltonian
with an interaction
\begin{equation}\label{inter}
H_I = J\sum_{k,k',M,M'}  f^{\dagger}_{M} c_{k,M} c^\dagger_{k',M'} f_{M'}
\end{equation}
with $J=V^2U/[\mu_f(\mu_f+U)]$.
This is a Kondo type\cite{kondo}
interaction and describes the coupling of the fluctuating $M$ state at the {\em Ce} site to the
 conduction band.
 Alternately we may write
 \begin{equation}\label{fint}
 \begin{split}
 H_I = J \sum_{\textbf{k}, \textbf{k}', \sigma, \sigma', M, M'} & <\textbf{k}' \sigma'|k' M'> <kM|\textbf{k} \sigma> \\
 & \times\ f^\dagger_M c_{\textbf{k} \sigma} c^\dagger_{\textbf{k}',\sigma'} f_{M'},
 \end{split}
 \end{equation}

\section{Kondo lattice model}
We now generalize the description of a single Ce impurity ion of the
previous section to a lattice of Ce ions. We first introduce
operators $f_{M,\textbf{R}}$ for the local moments at site
$\textbf{R}$ of the lattice. The generalization of the Kondo
interaction $H_I$ is clearly
\begin{equation}\begin{split}
H_K =J \
\sum_{\textbf{R}}\sum_{\textbf{k},\sigma,\textbf{k}',\sigma',M,M'}
&\langle  \textbf{k}',
\sigma' | k',M',\textbf{R} \rangle \langle k,M,\textbf{R}| \textbf{k},\sigma  \rangle\\
& \times \  f_{\textbf{R},M}^\dagger c_{\textbf{k},\sigma}
c^\dagger_{\textbf{k}',\sigma'}  f_{\textbf{R},M'},
\end{split}
\end{equation}
where $| k,M,\textbf{R} \rangle$ is a $c$-electron partial wave
centered at site $\textbf{R}$. We have
\begin{equation}
| k,M,\textbf{R} \rangle=e^{i\hat{\textbf{P}} \cdot \textbf{R}} |
k,M \rangle,
\end{equation}
where $\hat{\textbf{P}}$ is the momentum operator (generator of
translation) and $| k,M \rangle$ is a partial wave centered at the
origin. Thus we get:
\begin{equation}
\langle  \textbf{k}, \sigma | k,M,\textbf{R} \rangle=\langle
\textbf{k}, \sigma |e^{i\hat{\textbf{P}} \cdot \textbf{R}} | k,M
\rangle = e^{i \textbf{k}\cdot\textbf{R}}\langle  \textbf{k}, \sigma
| k,M \rangle,
\end{equation}
 since $\langle  \textbf{k}, \sigma |$ is momentum eigen state. With Fourier transforming the $c_{\textbf{k}}$ electrons back
 to real space ($ c_{\textbf{k},\sigma}=\sum_{\textbf{r}}
e^{i\textbf{k}\cdot \textbf{r}} c_{\textbf{r},\sigma} $) we get:
\begin{equation}\begin{split}
H_K =J \ \sum_{\textbf{R}} & \sum_{\textbf{r},\textbf{r'},M,M'}
  f_{M,\textbf{R}}^\dagger\left[ \sum_{\textbf{k},\sigma}\langle  k,M |
\textbf{k},\sigma \rangle  e^{i\textbf{k}\cdot
(\textbf{r}-\textbf{R})} c_{\textbf{r},\sigma}\right]
\\ & \left[ \sum_{\textbf{k}',\sigma'}\langle \textbf{k'},\sigma'|
k',M'\rangle  e^{-i\textbf{k}'\cdot (\textbf{r}'-\textbf{R})}
c^\dagger_{\textbf{r}',\sigma'}\right] f_{M',\textbf{R}}.
\end{split}
\end{equation}
It is convenient now to define real space operators
\begin{equation}
\Gamma_{\textbf{r},\textbf{R},M} =\sum_{\textbf{k},\sigma}\langle
k,M | \textbf{k},\sigma \rangle e^{i\textbf{k}\cdot
(\textbf{r}-\textbf{R})} c_{\textbf{r},\sigma},
\end{equation}
which are a
mixture of spin up and down electrons. In terms of these real-space operators,
the Kondo interaction assumes
simple form:
\begin{equation}
H_K =J\ \sum_{\textbf{R}} \sum_{\textbf{r},\textbf{r}'} \sum_{M,M'}
f_{M,\textbf{R}}^\dagger \Gamma_{\textbf{r},\textbf{R},M}
\Gamma_{\textbf{r}',\textbf{R},M'}^\dagger f_{M',\textbf{R}'}.
\end{equation}

The full Kondo lattice model then takes the form
\begin{eqnarray}
H & = & H_c + H_K \\
H_c & = & \sum_{\textbf{k}, \sigma} \varepsilon_{\textbf{k}}
c^{\dagger}_{\textbf{k},\sigma} c_{\textbf{k},\sigma},
\end{eqnarray}
together with the constraints
\begin{equation}\label{constr}
\sum_M f^{\dagger}_{M, \textbf{R}}f_{M, \textbf{R}} = 1,
\end{equation}
at each site $\textbf{R}$. Note that due to this constraint it is no
longer appropriate to think of the $f$-operators as describing
physical electrons. Rather at this stage they should be viewed as neutral fermions
that carry spin alone. As is well known this representation is redundant and introduces an
extra $U(1)$ gauge structure associated with the freedom to change the phase of $f$ independently at each site.

\section{Slave boson mean field theory}\label{slave}
We now discuss the Fermi liquid phases described by this Kondo
lattice model within the slave boson mean field approximation. In
simpler Kondo lattice models this technique correctly captures the
essential physics of the fermi liquid state\cite{millee}. In the
mean field we impose the constraint of Eqn. \ref{constr} on average
with a chemical potential $\mu_f$ for the $f$-fermions and replace
the Kondo interaction by a self-consistently determined
hybridization between the $c$ and $f$ operators. The mean field
Hamiltonian reads
\begin{equation}
\begin{split}
H_{MF} = & \sum_{\textbf{k} \sigma} \varepsilon_{\textbf{k}} c^{\dagger}_{\textbf{k}\sigma} c_{\textbf{k}\sigma} + \mu_f \sum_{M\textbf{R}} f^{\dagger}_{M, \textbf{R}}f_{M, \textbf{R}} \\
& + b \sum_{M\textbf{R}} \left(f^{\dagger}_{M\textbf{R}}
\sum_{\textbf{r}} \Gamma_{\textbf{r} \textbf{R} M} + h.c \right).
\end{split}
\end{equation}
The mean field parameters $\mu_f, b$ must be determined self-consistently through the equations
\begin{eqnarray}
  1 & = & \sum_M \langle f^{\dagger}_{M, \textbf{R}}f_{M, \textbf{R}} \rangle,   \\
\label{scm} b & = & J\langle \sum_M f^{\dagger}_{M \textbf{R}}
\sum_{\textbf{r}} \Gamma_{\textbf{r} \textbf{R} M} \rangle.
\end{eqnarray}
Note that we have chosen $b$ to be real in this mean field. Parenthetically we note that
a non-zero mean field
hybridization parameter $b$ should really be viewed as a Higgs condensate for the $U(1)$ gauge structure
introduced when we represent the spins in terms of the $f$-fields. In this Higgs phase the internal gauge charge of the $f$-fermions is screened by the condensate and the resulting screened gauge neutral object has the same quantum numbers as the electron. This structure of the low energy electrons manifests itself as a small electron quasiparticle weight at the heavy electron Fermi surface.

To diagonalize this mean field Hamiltonian we go to momentum space.
We write  $f_{M,\textbf{R}}=\sum_{\textbf{q}} e^{-i\textbf{q}\cdot
\textbf{R}} f_{M,\textbf{q}}$ and put in original form of $\Gamma$
operators in terms of $c$. The hybridization term then becomes
\begin{equation}\begin{split}
H_{MF}=b\sum_{\textbf{R},\textbf{r}} & \sum_{\textbf{q}}
e^{i\textbf{q}\cdot \textbf{R}} f^\dagger_{M,\textbf{q}}
\sum_{\textbf{k},\sigma} \langle
 k,M |\textbf{k},\sigma \rangle e^{i\textbf{k}\cdot (\textbf{r}-\textbf{R})}
c_{\textbf{r},\sigma} \\ & + h.c.\\
=b\sum_{\textbf{q},\textbf{k},\sigma} & \left(\sum_{\textbf{R}}
e^{i(\textbf{q}-\textbf{k})\cdot\textbf{R}}\right)f^\dagger_{M,\textbf{q}}\
 \langle k,M |\textbf{k},\sigma \rangle \\ & \times\left(\sum_{\textbf{r}}
e^{i\textbf{k}\cdot \textbf{r}} c_{\textbf{r},\sigma}\right)+
h.c.\\=b\sum_{\textbf{k}}&  \langle
 k,M |\textbf{k},\sigma \rangle \
f^\dagger_{M,\textbf{k}}\ c_{\textbf{k},\sigma} +h.c. \ .
\end{split}\end{equation}
Thus the momentum dependence of the hybridization is captured through
the $\langle \textbf{k},\sigma | k,M \rangle$ matrix element, which
we calculate in the Appendix.

We  define the four component field $\Psi_{\textbf{k}}$
\begin{equation}\nonumber
 \Psi_\textbf{k}= \left[
\begin{array}{ccc}
c_{\textbf{k}\uparrow} \\ c_{\textbf{k}\downarrow} \\
f_{k,1}\\f_{k,2}
\end{array} \right],
\end{equation}
in terms of which, the Hamiltonian becomes
\begin{equation}\label{evalue}
H_{MF}= \sum_{\textbf{k}} \Psi^\dagger_{\textbf{k}}\left[
\begin{array}{cccc} \varepsilon_{\textbf{k}} \textbf{I} & b\ M(\textbf{k}) \\
b\ M^\dagger(\textbf{k}) & \mu_f \textbf{I}
\end{array} \right]\Psi_{\textbf{k}}.
\end{equation}
Here $M(\textbf{k})$ is a $2 \times 2$ matrix given by
\begin{equation}
M(\textbf{k})=\left[
\begin{array}{ccc}
B(\textbf{k}) &
  A^*(\textbf{k}) \\ A(\textbf{k}) & -B^*(\textbf{k})
\end{array} \right].
\end{equation}
The functions $A(\textbf{k}), B(\textbf{k})$ are defined in the
Appendix.

Now we look for operators $\gamma_i(\textbf{k})$ that satisfy
$[H_{MF},\gamma^{\dagger}_i(\textbf{k})]=\lambda_i(\textbf{k})
\gamma^{\dagger}_i(\textbf{k})$, in term of which $H^{MF}$ is
diagonal:
\begin{equation}
H_{MF}=\sum_{\textbf{k},i} \lambda_i(\textbf{k})
\gamma^{\dagger}_i(\textbf{k}) \gamma_i(\textbf{k}).
\end{equation}
If we express $\gamma_i(\textbf{k})$ as:
\begin{equation}
\gamma_i(\textbf{k})=u^1_i(\textbf{k})
c_{\textbf{k}\uparrow}+u^2_i(\textbf{k})c_{\textbf{k}\downarrow}+u^3_i(\textbf{k})
f_{\textbf{k} 1}+u^4_i(\textbf{k})f_{\textbf{k} 2},
\end{equation}
where the coefficients $u^{\alpha }_{i} (\textbf{k})$ are determined through the eigenvalue equation:
\begin{equation}
\left[
\begin{array}{cccc} \varepsilon_{\textbf{k}} \textbf{I} & b\ M(\textbf{k}) \\
b\ M^\dagger(\textbf{k}) & \mu_f \textbf{I}
\end{array} \right]\left[
\begin{array}{ccc}
u_i^1(\textbf{k}) \\ u_i^2(\textbf{k}) \\
u_i^3(\textbf{k})\\u_i^4(\textbf{k})
\end{array} \right]=\lambda_i(\textbf{k}) \left[
\begin{array}{ccc}
u_i^1(\textbf{k}) \\ u_i^2(\textbf{k}) \\
u_i^3(\textbf{k})\\u_i^4(\textbf{k})
\end{array} \right]
\end{equation}
 From this we get the four eigenstates and the corresponding
dispersion of four bands:
 \begin{eqnarray}
 \nonumber\lambda_1(\textbf{k})&=&\frac{\varepsilon_\textbf{k}+\mu_f}{2}\\ \nonumber &-& \sqrt{\left(\frac{\varepsilon_\textbf{k}-\mu_f}{2}\right)^2+b^2(|A(\Omega_{\textbf{k}})|^2+|B(\Omega_{\textbf{k}})|^2)},\\
 \nonumber\lambda_2(\textbf{k})&=&\frac{\varepsilon_\textbf{k}+\mu_f}{2}\\ \nonumber &-& \sqrt{\left(\frac{\varepsilon_\textbf{k}-\mu_f}{2}\right)^2+b^2(|A(\Omega_{\textbf{k}})|^2+|B(\Omega_{\textbf{k}})|^2)},\\
\nonumber \lambda_3(\textbf{k})&=&\frac{\varepsilon_\textbf{k}+\mu_f}{2}\\ \nonumber &+& \sqrt{\left(\frac{\varepsilon_\textbf{k}-\mu_f}{2}\right)^2+b^2(|A(\Omega_{\textbf{k}})|^2+|B(\Omega_{\textbf{k}})|^2)},\\
\nonumber\lambda_4(\textbf{k})&=&\frac{\varepsilon_\textbf{k}+\mu_f}{2}.\\
\nonumber &+&
\sqrt{\left(\frac{\varepsilon_\textbf{k}-\mu_f}{2}\right)^2+b^2(|A(\Omega_{\textbf{k}})|^2+|B(\Omega_{\textbf{k}})|^2)}
\end{eqnarray}
At each $\textbf{k}$ obviously we have
$\lambda_1(\textbf{k})=\lambda_2(\textbf{k})\leq\lambda_3(\textbf{k})=\lambda_4(\textbf{k})$,
so we have two sets of doubly-degenerate bands. The degeneracy is a consequence of time reversal and inversion
symmetries which have been assumed in the original model.

\begin{figure}[htp]
\includegraphics[height=5cm]{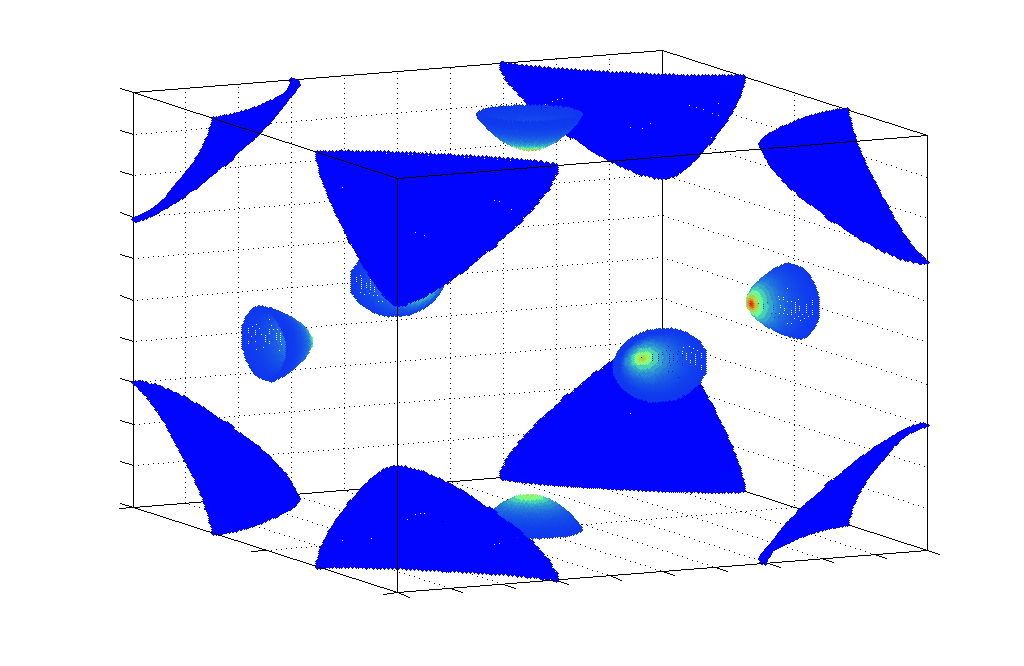}
\caption{$Z$ on Fermi surface. Red denotes larger $Z$ close to one
and blue denotes $Z$ close to zero. Red points are along
(0,0,1),(0,1,0) and (0,0,1) directions.} \label{fermisf}
\end{figure}

Let us assume that there are $n_c$ conduction electrons per unit
cell with $n_c <1$. Once combined with single $f$-fermion per unit
cell, we then need to fill these bands up to the Fermi energy to give
a total particle number of $1+n_c$ per unit
cell. Only only states in the lower bands $\lambda_1$ and $\lambda_2$ are
e filled, and the Fermi surface
always lives in these two bands. Clearly the Fermi surface is
large in that its volume counts both the conduction electrons and
the $f$-fermions. The shape of the Fermi surface corresponding to our
simple model is shown shown in Fig. \ref{fermisf}.

We note that the hybridization matrix $bM(\vec k)$ vanishes along
the $(100)$ and symmetry related directions (see Appendix). These
hybridization nodes lead to striking Fermi surface anisotropies as
discussed in detail below. For now we note that along these
`nodal' directions the Fermi surface coincides within the original {\em
small} conduction electron Fermi surface. To see this consider the
spectrum of the partially occupied band. It is
obvious that
$\lambda(\textbf{k})<\frac{\varepsilon_\textbf{k}+\mu_f}{2}-|\frac{\varepsilon_\textbf{k}-\mu_f}{2}|$.
On the other hand
$\frac{\varepsilon_\textbf{k}+\mu_f}{2}-|\frac{\varepsilon_\textbf{k}-\mu_f}{2}|=min\{\varepsilon_\textbf{k},\mu_f\}$
so that for all $\textbf{k}$, $\lambda(\textbf{k})\leq\mu_f$
(equality only holds for the points where $b(\textbf{k})=0$). So
for $n_c <1$, $E_f<\mu_f$. Now if we consider points on the fermi
surface where $b(\textbf{k})$ vanishes, again for such points
$\lambda(\textbf{k})$ is equal to either $\mu_f$ for points where
$\varepsilon_\textbf{k}>\mu_f$ and $\varepsilon_\textbf{k}$ for
points where $\varepsilon_\textbf{k}<\mu_f$. Since $E_f<\mu_f$ for
such points we have $\lambda(\textbf{k})=\varepsilon_\textbf{k}$;
so the Fermi surface coincides with the small Fermi sea of conduction
electrons at the points where $b(\textbf{k})=0$. Thus along
the nodal directions of the hybridization the quasiparticles at
the Fermi surface are almost entirely composed of
conduction-electrons.
However, on
moving away from these nodal directions the quasiparticles quickly
acquire an almost complete $f$-character  with a weak conduction
electron  admixture.

\section{Implications for photoemission experiments}
The anisotropic hybridization leads, as discussed below, to
anisotropic quasiparticle spectral weight. This can be probed by
angle resolved photoemission spectroscopy (ARPES). We begin with a
general discussion of the physical electron Green function in
Kondo lattice systems.

Let us start with Anderson model given in equation \ref{ande}. In
ARPES by interaction with a light beam electrons are extracted from
the sample. These electrons in principle could be extracted form any
of the two bands (see Equation \ref{ande}). However processes where
the $f$-occupation is changed cost large energy in the strong
correlation limit. On the other hand, processes where the removed
$f$-electron is replaced by tunneling of a $c$-electron into the
unoccupied $f$-site can occur at order($V/U)$ and will have matrix
elements in the low energy ``Kondo" subspace. To discuss this
physics let us start from a single Anderson impurity and consider
the operator $\psi_{\textbf{k},\sigma}$ that corresponds to
extracting a electron with momentum $\textbf{k}$ and spin $\sigma$
out of the sample:

\begin{equation}
\psi_{\textbf{k},\sigma}=c_{\textbf{k},\sigma}+\sum_M \langle
\textbf{k},\sigma | M \rangle f_M.
\end{equation}
In the strong correlation limit we need to perform the
Schrieffer-Wolff transformation for this operator. Below we use an
equivalent alternate procedure. We first consider the ground state.
In the limit of infinite $U$ it consists of half filled $f$ orbital
coexisting with a conductino band filled up to the Fermi energy:
we name this state $|g_0\rangle$. In the
limit of large but finite $U$  to first order in $V/U$ the
ground state becomes:
\begin{equation}
|g_1 \rangle=|g_0 \rangle+\sum_{n,M,\textbf{k},\sigma}|n\rangle \frac{\langle
n |V \left(f^\dagger_M
c_{\textbf{k},\sigma}+c_{\textbf{k},\sigma}^\dagger f_M \right)|g_0
\rangle}{U}
,
\end{equation}
where, for simplicity we have assumed that the energy to add, or
remove an f-electron is $U$.
Here $|n\rangle$ denote the first excited states. To a good
approximation, the state-vector in the second term is given by
written as $\left(f_M^\dagger
c_{\textbf{k},\sigma}+c_{\textbf{k},\sigma}^\dagger f_M
\right)|g_0\rangle$.
Now when we annihilate an electron by acting
with $\psi$ on $|g_1\rangle$ state, only final states which lie within the manifold of
states with single $f$-occupancy at the impurity, site will
contribute to the photoemission intensity at low energy. There are
two such states, one corresponding to the action of $c$ on the $|g_0
\rangle$ component of the ground state, and the other
corresponding to the action of $f$ on the second term in $\vert
g_{1}\rangle $ ({\em i.e} on the $f^\dagger c | g_0 \rangle$ term).
The net action of $\psi_{\textbf{k},\sigma}$ on $|g_0\rangle$ is then:

\begin{equation}\begin{split}
\psi_{\textbf{k},\sigma} \sim c_{\textbf{k},\sigma}+(V/U) &
\sum_{\textbf{k}',\sigma',M',M}\langle k', M' |\textbf{k}',\sigma'
\rangle \\ & \times \langle \textbf{k},\sigma | k,M\rangle \
f^\dagger_{M'} c_{\textbf{k}',\sigma'}f_M.
\end{split}\end{equation}

The first term corresponds to the knocking off an electron from
the $c$ band  and second term corresponds to the first order
process where an electron from an $f$ orbital is knocked off and
an electron from $c$ band replaces it.
 Now for a lattice of impurities, we should consider processes
where $f$ electrons from different sites are knocked out:

\begin{equation}\label{c-oper}\begin{split}
\psi_{\textbf{k},\sigma} \sim c_{\textbf{k},\sigma}+(V/U)
\sum_{\textbf{R}} & \sum_{\textbf{k}',\sigma',M',M}\langle k',
M',\textbf{R} |\textbf{k}',\sigma' \rangle   \\ & \times \langle
\textbf{k},\sigma | k,M,\textbf{R}\rangle f^\dagger_{M',\textbf{R}}
\  c_{\textbf{k}',\sigma'}f_{M,\textbf{R}}.
\end{split}\end{equation}

It is convenient to reexpress this in real space. The procedure is
the same we did in section \ref{slave}:
\begin{equation}\begin{split}
\sum_{\textbf{R}} \sum_{\textbf{k}',\sigma',M'} & \langle k',
M',\textbf{R} |\textbf{k}',\sigma' \rangle   \langle
\textbf{k},\sigma | k,M,\textbf{R}\rangle f^\dagger_{M',\textbf{R}}
c_{\textbf{k}',\sigma'}f_{M,\textbf{R}}\\ =\sum_{\textbf{R}}
\sum_{\textbf{k}',\sigma',M'} & \langle k', M' |\textbf{k}',\sigma'
\rangle e^{i\textbf{k}'.(\textbf{r}-\textbf{R})}
f^\dagger_{M',\textbf{R}}c_{\textbf{r},\sigma'} \\ &\ \ \ \ \ \ \
\langle \textbf{k},\sigma | k,M\rangle e^{i \textbf{k}.\textbf{R}}
f_{M,\textbf{R}} \\ =\sum_{\textbf{R}} \sum_{M'} &
f^\dagger_{M',\textbf{R}} \left[\sum_{\textbf{k}',\sigma}\langle k',
M' |\textbf{k}',\sigma' \rangle
e^{i\textbf{k}'.(\textbf{r}-\textbf{R})}c_{\textbf{k}',\sigma'}\right]
\\ & \sum_M e^{i \textbf{k}.\textbf{R}}
f_{M,\textbf{R}} \\
=\sum_{\textbf{R}}\sum_{M'} & f^\dagger_{M',\textbf{R}}
\Gamma_{\textbf{r},\textbf{R},M'} \sum_M \langle \textbf{k},\sigma |
k,M\rangle f_{M,\textbf{R}}.
\end{split}\end{equation}
Within the slave boson mean field approximation  we replace the
product $f^\dagger c$ (or equivalently $f^\dagger \Gamma$)  term in
the second term by its average to get:
\begin{equation}\label{arpo}
\psi_{\textbf{k},\sigma} \sim c_{\textbf{k},\sigma}+(b/V)
\sum_{M}\langle \textbf{k},\sigma | k,M\rangle f_M. \end{equation}

The ARPES intensity may now be calculated from  the Greens
function of this $\psi$ operator. Its trace is given by
\begin{equation}\label{tgrn}\begin{split}
Tr&\left[\textit{G}_{\sigma,\
\sigma'}(k,i\omega_\nu)\right]=\int_0^\beta d\tau
e^{i\omega_\nu\tau}\\ &\langle T_\tau \left[\psi_{\uparrow}(k,\tau)
\psi_{\uparrow}^\dag(k,0)+\psi_{\downarrow}(k,\tau)
\psi_{\downarrow}^\dag(k,0)\right]\rangle,
\end{split}\end{equation}
where the expectation value is taken in the ground state. From
equation \ref{arpo} it is obvious that this green function consists
four different terms. For this calculation, we need to have
$c_\sigma$ and $f_M$ operators, in term of $\gamma$ operators. To
make this calculation more transparent, it is useful to introduce
the unitary matrix $U$ as:
\begin{equation}
U=\left[
\begin{array}{cccc} u^1_1 & u^2_1 & u^3_1 & u^4_1 \\
 u^1_2 & u^2_2 & u^3_2 & u^4_2\\ u^1_3 & u^2_3 & u^3_3 & u^4_3
\\ u^1_4 & u^2_4 & u^3_4 & u^4_4 \end{array} \right],
\end{equation}
where:
\begin{equation}
\left[
\begin{array}{ccc} \gamma^1 \\ \gamma^2 \\ \gamma^3 \\ \gamma^4
\end{array} \right] = U \left[
\begin{array}{ccc}
c_{\uparrow} \\ c_{\downarrow} \\
f_{1}\\f_{2}
\end{array} \right].
\end{equation}
Here $\textbf{k}$ index is suppressed for notational
convenience. Inverting we get
\begin{eqnarray}
c^\dagger_\uparrow &=& u^{1}_1 \gamma^\dagger_1 + u^{1}_2
\gamma^\dagger_2+ u^{1}_3 \gamma^\dagger_3+u^{1}_4 \gamma^\dagger_4 ,\\
c^\dagger_\downarrow &=& u^{2}_1 \gamma^\dagger_1 + u^{2}_2
\gamma^\dagger_2+ u^{2}_3 \gamma^\dagger_3+u^{2}_4 \gamma^\dagger_4,
\\ f^\dagger_1 &=& u^{3}_1 \gamma^\dagger_1 + u^{3}_2
\gamma^\dagger_2+ u^{3}_3 \gamma^\dagger_3+u^{3}_4
\gamma^\dagger_4,\\ f^\dagger_2 &=& u^{4}_1 \gamma^\dagger_1 +
u^{4}_2 \gamma^\dagger_2+ u^{4}_3 \gamma^\dagger_3+u^{4}_4
\gamma^\dagger_4.
\end{eqnarray}

Using this result, we can expand imaginary part of the trace of the Green
function to obtain the zero temperature spectral function. This has
four
terms corresponding to the operator combinations
$cc^\dagger$,$ff^\dagger$, $fc^\dagger$ and
$cf^\dagger$. Let us calculate them one by one. The
$cc^\dagger$ term is:
\begin{equation}\begin{split}
A_{cc}(\textbf{k},\omega)=
&(|u^1_1(\textbf{k})|^2+|u^2_1(\textbf{k})|^2)
\ \delta(\lambda_1(\textbf{k})-\omega)\\
& +(|u^1_2(\textbf{k})|^2+|u^2_2(\textbf{k})|^2) \
\delta(\lambda_2(\textbf{k})-\omega).
\end{split}\end{equation}

We then get the following form for the quasi-particle residue
on the fermi surface:

\begin{equation}\label{z}\begin{split}
Z_{cc}&(\textbf{k}|
\lambda_2(\textbf{k})=E_f)=|u^1_2(\textbf{k})|^2+|u^2_2(\textbf{k})|^2
=
\\ &
\frac{b(\textbf{k})^2}{b(\textbf{k})^2+\left(\frac{\varepsilon(\textbf{k})-\mu_f}{2}+
\sqrt{(\frac{\varepsilon(\textbf{k})-\mu_f}{2})^2+b(\textbf{k})^2}\right)^2}
\end{split}
\end{equation}

Now for $ff$ term (noting $u_1^3=u_2^4=0$) we have:
\begin{equation}\begin{split}
A_{ff}(\textbf{k},\omega)=
&\left(\frac{b(\textbf{k})}{V}\right)^2|u^1_4(\textbf{k})|^2
\ \delta(\lambda_1(\textbf{k})-\omega)\\
& +\left(\frac{b(\textbf{k})}{V}\right)^2|u_2^3(\textbf{k})|^2 \
\delta(\lambda_2(\textbf{k})-\omega).
\end{split}\end{equation}

This gives the residue:
\begin{equation}\label{z}\begin{split}
Z_{ff}&(\textbf{k}| \lambda_2(\textbf{k})=E_f)=\\
&\frac{\left(\frac{\varepsilon(\textbf{k})-\mu_f}{2}+
\sqrt{(\frac{\varepsilon(\textbf{k})-\mu_f}{2})^2+b(\textbf{k})^2}\right)^2}{b(\textbf{k})^2+\left(\frac{\varepsilon(\textbf{k})-\mu_f}{2}+
\sqrt{(\frac{\varepsilon(\textbf{k})-\mu_f}{2})^2+b(\textbf{k})^2}\right)^2}\left(\frac{b(\textbf{k})}{V}\right)^2.
\end{split}
\end{equation}
The last contribution will be:
\begin{equation}\label{z}\begin{split}
Z_{cf}&(\textbf{k}| \lambda_2(\textbf{k})=E_f)=\\
&-\frac{2(b^2/V)\left(\frac{\varepsilon(\textbf{k})-\mu_f}{2}+
\sqrt{(\frac{\varepsilon(\textbf{k})-\mu_f}{2})^2+b(\textbf{k})^2}\right)}{b(\textbf{k})^2+\left(\frac{\varepsilon(\textbf{k})-\mu_f}{2}+
\sqrt{(\frac{\varepsilon(\textbf{k})-\mu_f}{2})^2+b(\textbf{k})^2}\right)^2}\\
& \ \ \ \
\times\Re\left(A^2(\Omega_{\textbf{k}})+B^2(\Omega_{\textbf{k}})\right),
\end{split}
\end{equation}
where $b(\textbf{k})=b
\sqrt{|A(\Omega_{\textbf{k}})|^2+|B(\Omega_{\textbf{k}})|^2}$. In
Fig. \ref{fermisf} we have also indicated the total residue
$Z_{total}$ which is the sum of these three contributions.

Using the fact that $|b(\textbf{k})|$ is small, we can investigate
the behavior of $Z_{total}$ at least for the points where
$|b(\textbf{k})| \ll |\varepsilon(\textbf{k})-\mu_f|$. For such
points we see that whenever $\epsilon_\textbf{k} > \mu_f$ the
dominant term (of order $b^2/V^2$) is $Z_{ff}$ and it varies since
$b(\textbf{k})$ is angle dependent. On the other hand, when
$\epsilon_\textbf{k} < \mu_f$, the dominant contribution is
$Z_{cc}$ which is of order one. This information could be
summarized in the following form:
\begin{equation}\begin{split}
Z&(\textbf{k}| \lambda_2(\textbf{k})=E_f) =
\\ &
b^2\
\frac{h(\textbf{k})}{\left(\varepsilon(\textbf{k})-\mu_f\right)^2}\Theta\left(\varepsilon(\textbf{k})-\mu_f\right)+
\Theta\left(\mu_f-\varepsilon(\textbf{k})\right).
\end{split}
\end{equation}
A key result of this calculation is that for the points where
$\varepsilon(\textbf{k})>\mu_f$, $Z$ is small and of order
$\frac{b(\textbf{k})^2}{\left(\varepsilon(\textbf{k})-\mu_f\right)^2}$;
this quantity has varies by about $20\%$ due to the  angle dependent
$b(\textbf{k})$. But for the points where $\mu_f
> \varepsilon(\textbf{k})$, the quasi-particle residue will be of
order one and will exhibit no strong variations.
The small region in the middle of Fermi
surface in Fig. \ref{fermisf} with $Z \sim 1$
corresponds is these points. These regions are centered along
$(100)$ and symmetry related directions. As discussed in the
previous section the hybridization matrix has nodes in these
special directions and the corresponding quasiparticles are
essentially conduction electrons with
$Z \sim 1$.
On the other hand further away from these nodal
directions the quasiparticles develop $f$-character and
$Z \sim o(b^2)$ along these other directions.

There is thus a dramatic anisotropy in $Z$ on moving around the
Fermi surface. We note that ARPES experiments will
naturally be able to resolve the quasiparticle peak along high-$Z$
directions. However a low resolution ARPES study may well not
be able to resolve the small-$Z$ quasiparticles at all and may
incorrectly conclude that the fermi surface consists only of finite
open ended pieces.

\section{Momentum dependent effective mass}

It is well known that the effective mass $m^*$ in a heavy fermion system can be very anisotropic on the Fermi surface. How do these anisotropies correlate with the anisotropic $Z$?  It is precisely
the combination $Z \times m^*$ that determines the {\em tunneling} density of
states. It is therefore also interesting to look at
$m^*(\textbf{k})$ variations over the Fermi surface. The effective
mass can be calculated by taking the second derivative of energy
with respect to momentum in direction perpendicular to the Fermi
surface i.e. $\frac{\partial^2 \lambda_2(\textbf{k})}{\partial
k_\bot^2}$:
\begin{equation}\begin{split}
1/(m^*(\textbf{k}))=  1/2m_e^* &
\left[1-\frac{\frac{\varepsilon_{\textbf{k}}-\mu_f}{2}}{\sqrt{\left(\frac{\varepsilon_{\textbf{k}}-\mu_f}{2}\right)^2+b(\textbf{k})^2}}\right.\\
& \left. + \frac{b^2(\varepsilon_{\textbf{k}}-\mu_f)
f(\textbf{k})}{\left(\left(\frac{\varepsilon_{\textbf{k}}-\mu_f}{2}\right)^2+b(\textbf{k})^2\right)^{3/2}}\right]
\\ \approx\ & 1/m_e^*
\left[\Theta\left(\mu_f-\varepsilon(\textbf{k})\right)+
\frac{b^2}{(\varepsilon_{\textbf{k}}-\mu_f)^2} g(\textbf{k})\right]
\end{split}\end{equation}
where $m_e^*$ is the free electron effective mass
($1/m_e^*=\frac{\partial^2 \varepsilon_{\textbf{k}} }{\partial
k_\perp^2}$), and in the last step we used the approximation
$|\varepsilon_{\textbf{k}}-\mu_f|\gg b(\textbf{k})|$.
$f(\textbf{k})$ and $g(\textbf{k})$ are dimensionless functions of
$\textbf{k}$ where $g(\textbf{k})= 2\
sign(\varepsilon_{\textbf{k}}-\mu_f) \left(4
f(\textbf{k})+|A(\Omega_{\textbf{k}})|^2+|B(\Omega_{\textbf{k}})|^2
\right)$ (numerical calculations show no $\textbf{k}$ point where
$g(\textbf{k})$ vanishes). Inverting this we get:
\begin{equation}
m^*(\textbf{k})\approx m_e^*
\left[\Theta\left(\mu_f-\varepsilon(\textbf{k})\right)+\Theta\left(\varepsilon(\textbf{k})-\mu_f\right)\frac{(\varepsilon_{\textbf{k}}-\mu_f)^2}{g(\textbf{k})b^2}\right]
\end{equation}
We see a similar behavior with $Z(\textbf{k})$. Again for points
with $\varepsilon_{\textbf{k}}>\mu_f$ we have quasiparticles with
large effective mass, but for $\varepsilon_{\textbf{k}}<\mu_f$
quasiparticles are free electron types and have effective mass
corresponding to small, conduction electrons effective mass. We
see that we have large effective mass in the points where $Z$ is
small. So indeed variations of effective mass are correlated with
variations of $1/Z$.  The approximate invariance of the product $Z
({\bf k} )m^{*} (k)$ is a  momentum-space variant of Langreth  theorem, which
states that  the single particle density of states in the Anderson impurity model is
an adiabatic invariant, independent of the strength of the
interaction.
 
 This is interesting since it shows us that
the strong angle dependent anisotropy does not apparently have large
observable consequence on ordinary tunneling measurements. However it may possibly show up in the amplitude of the Friedel oscillations of the tunneling conductance around an impurity, and may therefore be accessible through Fourier transform scanning tunneling spectroscopy.


\section{Underdoped cuprates: Pseudogaps and Fermi arcs in a large Fermi surface metal? }\label{cuprates}
We now compare the phenomena described with
observations on the normal state of the cuprate materials. As
discussed above in the heavy fermion context there are portions of
the Fermi surface where $Z \sim o(1)$, and ARPES experiments may
conclude that the Fermi surface consists of open ended pieces. This
is strongly reminiscent of the Fermi arc phenomena reported by ARPES
in  the pseudogap regime of the underdoped cuprates. It is tempting therefore to imagine that a similar
mechanism is operational in the cuprates. More specifically is it
possible that the underdoped cuprates actually have a large
band-structure-like Fermi surface but the $Z$ is $o(1)$ only along
the observed Fermi arcs and becomes very small away from it so that
those portions are not easily observed? The antinodal pseudogap itself must then be associated with a gap in the incoherent part of the electron spectrum with the gapless coherent part not resolved due to the smallness of $Z$.

In considering this question we first observe that in the heavy
fermion system the smallness of $Z$ goes hand-in-hand with the
largeness of effective mass. More generally the effective mass is not
directly related to $Z$  (it is only in cases where the electron self energy is momentum independent that
$Z$ determines the mass renormalization).   So
phenomenologically we need to first suppose that the small $Z$
antinodal regions do not have mass enhancement. Such a Fermi liquid
state for the pseudogap regime has some attractive features.  Consider
first the gapless Fermi arcs. Several popular theories attempt to view
the arcs as part of a true Fermi surface which consists of small
closed hole pockets whose {\em back} portions are not observed in
ARPES due to a small $Z$. However, the observed Fermi arc coincides
with band structure Fermi surface and shows no tendency to bend away
into a closed hole pocket. In contrast in the state discussed above
the true Fermi surface is simply the band structure one but the
antinodal sections would be unobservable due to a small $Z$.

Consider next recent observations of quantum oscillations at high
fields and low temperatures in some underdoped
cuprates\cite{taill1,harrison}. The oscillation frequency seems
consistent with a small Fermi pocket. A key issue is to reconcile this
with the Fermi arcs reported in photoemission, and a few different
ideas have been proposed\cite{acl,patrickrev}.  An interesting feature
of the high field experiments is a negative Hall constant which has
been interpreted as evidence for an {\em electron}
pocket\cite{taill2}. Recently Millis and Norman\cite{milnorm} have
proposed that the oscillations and negative Hall constant should be
with a $1/8$th filling antiphase stripe order, which folds
the band structure  Fermi surface to create a pocket. One issue with the
proposal is that the electron pocket is near the edges of the full
Brillouin zone - precisely the region where a big pseudogap is seen by
ARPES in zero field in the normal state above $T_c$. For the theory of
Ref. \onlinecite{milnorm} to apply it is apparently necessary that the
$60 T$ fields used in the quantum oscillation experiment wipe out the
pseudogap\cite{paleec}. This may seem unnatural but is not prohibited. This
difficulty is overcome in the large Fermi surface pseudogap envisaged
in this section. A low temperature $1/8$ antiphase stripe instability
arising from that state will have retain all the same transport
properties as that in the theory of Ref. \onlinecite{milnorm}. This is
because the smallness of $Z$ does not affect transport phenomena. On
the other hand the ARPES pseudogap (which in this state is the gap of
the incoherent part of the spectrum) will survive intact. Thus this
kind of large Fermi surface state provides a possible route to a
reconciliation between the quantum oscillation and ARPES experiments.

However a number of difficulties exist with the idea that the
pseudogap state has a large Fermi surface state with strong angle
dependent $Z$. First, the density of states as measured by
thermodynamic measurements actually decreases on entering the
pseudogap state by cooling. This requires that the
effective mass at the antinodal regions is {\em suppressed} (rather
than enhanced) in the pseudogap state which is rather
unnatural. Besides such behavior should signal an increase in the
Drude weight in optical transport in the pseudogap state which is not
seen. Finally this is also inconsistent with the scaling of the
superfluid density with the density of doped holes.

In light of these difficulties it seems unlikely that a Fermi liquid
state with a large Fermi surface of the kind discussed here is a
serious candidate for the pseudogap state. These difficulties may
perhaps be overcome by a non-Fermi liquid version which retains the
large Fermi surface and the strong variation of the low energy
spectral density. However a description of such a state does not
currently exist.

\section{Discussion}
The most interesting conclusion from this work is the possibility of
large variations in the quasiparticle weight (and concomitantly the
effective mass) on moving around the Fermi surface. This
variation is related to the internal orbital structure of the Kondo
resonance, derived from the $f$-symmetry of the orbitals occupied by the
local moments.  In the hybridization mean field theory the most dramatic variation occurs when there are
`hybridization nodes', {\em i.e} directions along which the hybridization vanishes.
We demonstrated the possibility of such nodes in a simple model of a $Ce$-based cubic heavy fermion system.  Hybridization nodes
lead to the possibility that some
portions of the large Fermi surface are actually contained within
the original small Fermi surface of the conduction electrons. In
those regions the quasiparticles essentially have $c$-electron
character with very little admixture to the $f$-fermions. The
quasiparticle weight is correspondingly large (of order $1$). The
opposite is true in other portions where the quasiparticles mostly
have $f$-character and have small $Z$. This then leads to a strong
angle dependence of the quasiparticle weight.

Real heavy electron materials have much more complicated band
structures than in the simplified model considered here. Nevertheless
there exists in general the possibility of hybridization nodes which
will greatly affect their low temperature physics.  Consider for
instance heavy electron superconductivity. At least in some cases the
superconductivity may be driven by formation of singlet bonds between
neighbouring local moments due to RKKY interactions. In combination
with Kondo hybridization this leads to superconductivity. Formally the
singlet formation may be described as $<ff>$ pairing while the Kondo
hybridization has non-zero $<c^\dagger f>$. This then leads to
non-zero $<cc>$, {\em i.e} superconducting
order\cite{andreicol,voj}. If the hybridization has nodes then this
will lead to extra nodes in the physical superconducting order
parameter over and above any nodes inherited from the singlet bond
$<ff>$ amplitude\cite{dzero}.

The large variation of the $Z$ also has potential implications for
current thinking on the nature of the quantum critical point between
the heavy Fermi liquid and the antiferromagnetic metal. It has been
suggested that this transition is accompanied by the loss of Kondo
screening resulting in a reconstruction of the Fermi
surface\cite{colq,svs,qsi}. Such a reconstruction presumably
requires $Z$ to vanish through out the large Fermi surface on
approaching the transition from the paramagnetic side. For a
discussion on $Z$ vanishing at the heavy Fermion quantum critical
points see [\onlinecite{senan}]. The variation of $Z$ described in
this paper raises the question of whether the manner in which $Z$
vanishes also varies around the Fermi surface.

We also explored the possibility that the pseudogap state of the underdoped cuprates may be
a large fermi surface Fermi liquid state with a strongly angle dependent $Z$. While such a picture
has some very appealing features it has enough difficulties with experiments that it is unlikely to directly be a
relevant theory of the pseudogap state.

\section*{Acknowledgments}
We thank A.J. Millis, P.A. Lee, and L. Taillefer for useful discussions.
TS was supported by an award from The Research Corporation for which
he is grateful. PC is supported by NSF grant DMR 0605935.

\begin{appendix}
\section{Calculation of matrix element}
To calculate $\langle \textbf{k},\sigma | k,M \rangle$, we use the
known overlap of $| \textbf{k},\sigma \rangle$ and $| k,J_z
\rangle$\cite{bal} for $l=3$:
\begin{equation}\begin{split}
\langle \textbf{k},\sigma |& k,J_z \rangle= \\ &
4\pi\left[\alpha_{J_z} Y_3^{J_z+\frac{1}{2}}(\Omega_{\textbf{k}}) \
\delta_{\sigma,-\frac{1}{2}} +\beta_{J_z}
Y_3^{J_z-\frac{1}{2}}(\Omega_{\textbf{k}}) \
\delta_{\sigma,\frac{1}{2}}\right]
\end{split}
\end{equation}
where $Y_l^m(\Omega_{\textbf{k}})$ are associated Legender functions
and $\alpha_{J_z}=[(7+2J_z)/14]^{1/2}$ and
$\beta_{J_z}=[(7-2J_z)/14]^{1/2}$ are Clebsh-Gordan
coefficients\cite{bal}. Now using the forms given in \ref{egn} we
get the following for the two orbital states:
\begin{eqnarray}
\langle \textbf{k},\sigma | \textbf{1} \rangle &=&
\frac{1}{\sqrt{6}}\
\left[\frac{1}{\sqrt{7}}Y_3^{-2}(\Omega_{\textbf{k}}) \
\delta_{\sigma,-\frac{1}{2}}+\sqrt{\frac{6}{7}}
Y_3^{-3}(\Omega_{\textbf{k}}) \ \delta_{\sigma,\frac{1}{2}}\right] \nonumber \\
&-& \sqrt{\frac{5}{6}}\
\left[\sqrt{\frac{5}{7}}Y_3^{2}(\Omega_{\textbf{k}}) \
\delta_{\sigma,-\frac{1}{2}}+\sqrt{\frac{2}{7}}
Y_3^{1}(\Omega_{\textbf{k}}) \ \delta_{\sigma,\frac{1}{2}}\right]
\nonumber \\
\langle \textbf{k},\sigma | \textbf{2} \rangle &=&
\frac{1}{\sqrt{6}}\
\left[\sqrt{\frac{6}{7}}Y_3^{3}(\Omega_{\textbf{k}}) \
\delta_{\sigma,-\frac{1}{2}}+\frac{1}{\sqrt{7}}
Y_3^{2}(\Omega_{\textbf{k}}) \ \delta_{\sigma,\frac{1}{2}}\right] \nonumber \\
&-& \sqrt{\frac{5}{6}}\
\left[\sqrt{\frac{2}{7}}Y_3^{-1}(\Omega_{\textbf{k}}) \
\delta_{\sigma,-\frac{1}{2}}+\sqrt{\frac{5}{7}}
Y_3^{-2}(\Omega_{\textbf{k}}) \ \delta_{\sigma,\frac{1}{2}}\right]
\nonumber
\end{eqnarray}
It is more convenient to work with a simplified version of these
relations as:
\begin{eqnarray}
\langle \textbf{k},\sigma | \textbf{1} \rangle &=&
\frac{1}{\sqrt{42}}\left[Y_3^{-2}(\Omega_{\textbf{k}})-5\
Y_3^{2}(\Omega_{\textbf{k}})\right] \delta_{\sigma,-\frac{1}{2}} \nonumber \\
&&+\frac{1}{\sqrt{42}}\left[\sqrt{6}Y_3^{-3}(\Omega_{\textbf{k}})-\sqrt{10}\
Y_3^{1}(\Omega_{\textbf{k}})\right]
\delta_{\sigma,\frac{1}{2}}\nonumber \\
\langle \textbf{k},\sigma | \textbf{2} \rangle &=&
\frac{1}{\sqrt{42}}\left[\sqrt{6}Y_3^{3}(\Omega_{\textbf{k}})-\sqrt{10}\
Y_3^{-1}(\Omega_{\textbf{k}})\right]\delta_{\sigma,-\frac{1}{2}} \nonumber \\
&&+\frac{1}{\sqrt{42}}\left[Y_3^{2}(\Omega_{\textbf{k}})-5\
Y_3^{-2}(\Omega_{\textbf{k}})\right]
\delta_{\sigma,\frac{1}{2}}\nonumber
\end{eqnarray}
If we introduce new functions $A(\Omega_{\textbf{k}})$ and
$B(\Omega_{\textbf{k}})$:
\begin{eqnarray}
\langle \textbf{k},\sigma | \textbf{1} \rangle &=&
A(\Omega_{\textbf{k}}) \ \delta_{\sigma,-\frac{1}{2}} +
B(\Omega_{\textbf{k}})
\ \delta_{\sigma,\frac{1}{2}} \nonumber\\
\langle \textbf{k},\sigma | \textbf{2} \rangle &=&
-B^*(\Omega_{\textbf{k}}) \ \delta_{\sigma,-\frac{1}{2}} +
A^*(\Omega_{\textbf{k}}) \ \delta_{\sigma,\frac{1}{2}}
\end{eqnarray}
where $A(\Omega_{\textbf{k}})= \frac{4\pi
}{\sqrt{42}}\left[Y_3^{-2}(\Omega_{\textbf{k}})-5\
Y_3^{2}(\Omega_{\textbf{k}})\right]$ and
$B(\Omega_{\textbf{k}})=\frac{4\pi
}{\sqrt{42}}\left[\sqrt{6}Y_3^{-3}(\Omega_{\textbf{k}})-\sqrt{10}\
Y_3^{1}(\Omega_{\textbf{k}})\right]$.
\end{appendix}
\bibliography{rkondo7pc}
\end{document}